\documentclass[10pt,twocolumn,showpacs,preprintnumbers,amsmath,amssymb,aps,prl,superscriptaddress,longbibliography]{revtex4-2}
\usepackage{appendix}
\usepackage{bbm}
\usepackage{mathrsfs}
\usepackage{graphicx}
\usepackage{dcolumn}
\usepackage{bm}
\usepackage{braket}
\usepackage{amsmath}
\usepackage{amsfonts}
\usepackage[dvipsnames]{xcolor}
\usepackage[colorlinks=true,linkcolor=Blue,urlcolor=BlueViolet,citecolor=BlueViolet]{hyperref}
\usepackage{natbib}
\usepackage{multirow}
\usepackage{floatrow}
\usepackage{hyperref}

\begin{document}

\title{Intrinsic antiferromagnetic topological insulator and axion state in V$_2$WS$_4$}
\author{Yadong Jiang}
\affiliation{State Key Laboratory of Surface Physics and Department of Physics, Fudan University, Shanghai 200433, China}
\affiliation{Shanghai Research Center for Quantum Sciences, Shanghai 201315, China}
\author{Huan Wang}
\affiliation{State Key Laboratory of Surface Physics and Department of Physics, Fudan University, Shanghai 200433, China}
\affiliation{Shanghai Research Center for Quantum Sciences, Shanghai 201315, China}
\author{Kejie Bao}
\affiliation{State Key Laboratory of Surface Physics and Department of Physics, Fudan University, Shanghai 200433, China}
\affiliation{Shanghai Research Center for Quantum Sciences, Shanghai 201315, China}
\author{Jing Wang}
\thanks{wjingphys@fudan.edu.cn}
\affiliation{State Key Laboratory of Surface Physics and Department of Physics, Fudan University, Shanghai 200433, China}
\affiliation{Shanghai Research Center for Quantum Sciences, Shanghai 201315, China}
\affiliation{Institute for Nanoelectronic Devices and Quantum Computing, Fudan University, Shanghai 200433, China}
\affiliation{Hefei National Laboratory, Hefei 230088, China}
\begin{abstract}
Intrinsic magnetic topological insulators offers an ideal platform to explore exotic topological phenomena, such as axion electrodynamics, quantum anomalous Hall (QAH) effect and Majorana edge modes. However, these emerging new physical effects have rarely been experimentally observed due to the limited choice of suitable materials. Here, we predict the van der Waals layered V$_2$WS$_4$ and its related materials show intralayer ferromagnetic and interlayer antiferromagnetic exchange interactions. We find extremely rich magnetic topological states in V$_2$WS$_4$, including an antiferromagnetic topological insulator, the axion state with the long-sought quantized topological magnetoelectric effect, three-dimensional QAH state, as well as a collection of QAH insulators and intrinsic axion insulators in odd- and even-layer films, respectively. Remarkably, the N\'eel temperature of V$_2$WS$_4$ is predicted to be much higher than that of MnBi$_2$Te$_4$. These interesting predictions, if realized experimentally, could greatly promote the topological quantum physics research and application.
\end{abstract}

\date{\today}

\maketitle

The discovery of intrinsic magnetic topological insulators (TIs)~\cite{hasan2010,qi2011,tokura2019,wang2017c,bernevig2022,chang2023} has opened new avenues for realizing a wide range of exotic topological phenomena through the time-reversal-breaking topological surface states~\cite{qi2008,essin2009,qi2009b,chen2010,li2010,wang2016a,yu2010,chang2013b,checkelsky2014,kou2014,bestwick2015,chang2015,mong2010,wan2011,xu2011,nomura2011,wang2015b,morimoto2015,mogi2017,mogi2017a,grauer2017,xiao2018,fu2008,fu2009a,qi2010b,wang2015c,lian2018b,wang2018}. A paradigm example is the realization of the quantum anomalous Hall (QAH) effect and axion insulator in few layer MnBi$_2$Te$_4$~\cite{deng2020,zhang2019,li2019,otrokov2019,gong2019,liu2020,deng2021}. Despite progress, experimental studies of magnetic topological states lag significantly behind their non-magnetic counterparts due to the limited availability of magnetic TI materials. While hundreds of intrinsic magnetic topological materials have been identified by symmetry-based analysis~\cite{bradlyn2017,slager2017,elcoro2021,po2017,watanabe2018,po2020}, \emph{ab initio} calculations~\cite{xu2020} and machine learning approaches~\cite{xu2024}, the vast majority are semimetals. 
To date, MnBi$_2$Te$_4$ remains the only experimentally confirmed intrinsic antiferromagnetic (AFM) TI~\cite{otrokov2019,gong2019}. 
However, MnBi$_2$Te$_4$ has a relatively low N\'eel temperature, and its complex magnetic structure, coupled with imperfect sample quality, has hindered direct observation of the exchange gap in Dirac surface states using spectroscopy measurements~\cite{gong2019,hao2019,lihang2019,chen2019}. Thus, the search for realistic intrinsic magnetic TIs, preferably with higher magnetic ordering temperatures and large gaps, has become an important goal in topological material research. In this context, the class of V$_2$WS$_4$ materials predicted in this paper offer a promising solution. These materials feature high N\'eel temperatures and provide an ideal platform for exploring emergent magnetic topological phenomena, such as AFM TIs, the topological axion state with topological magnetoelectric effect (TME), and the QAH effect in both two- and three-dimensional (3D) systems, and so on.

\begin{figure}[t]
\begin{center}
\includegraphics[width=3.4in, clip=true]{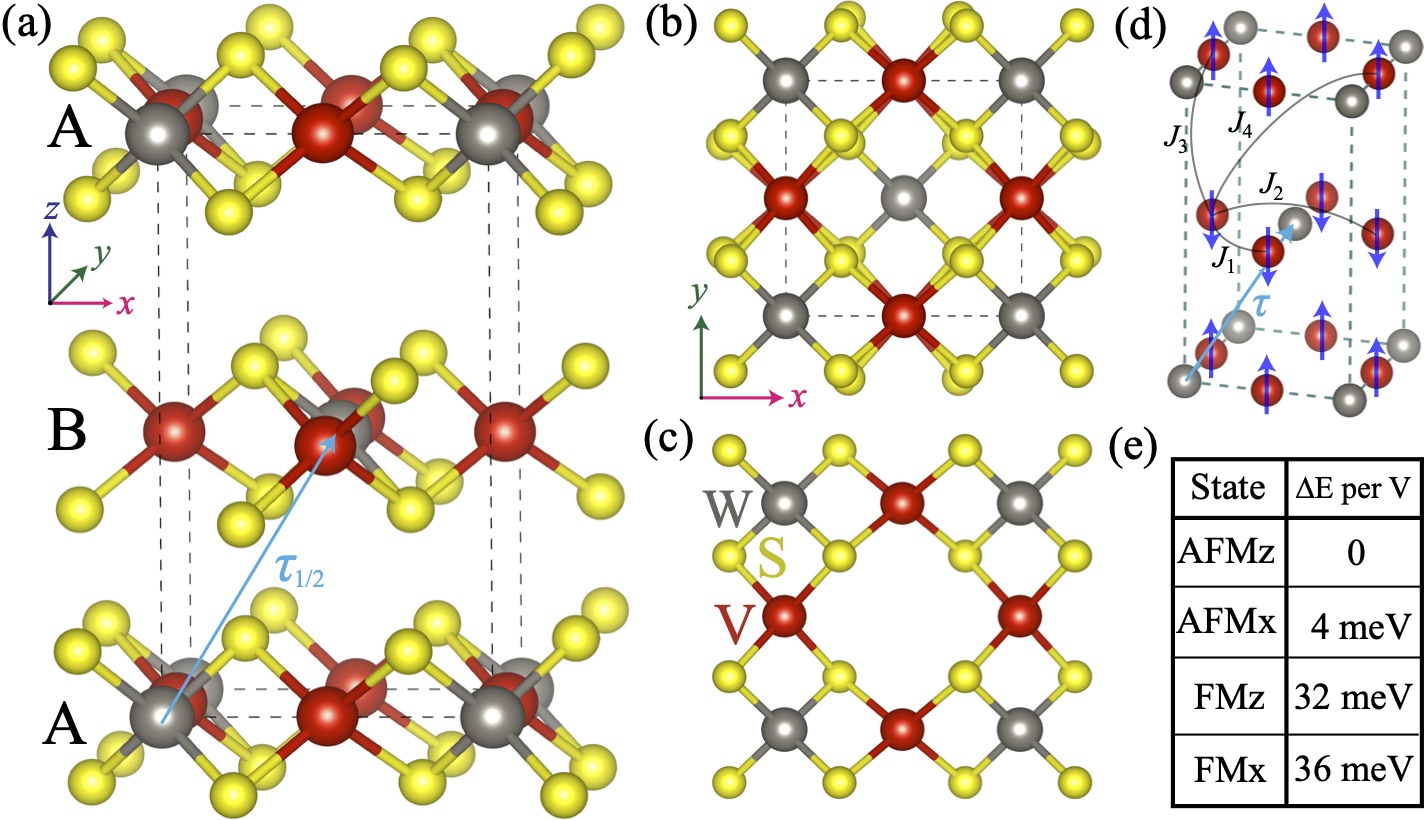}
\end{center}
\caption{Crystal and magnetic structures. (a) The unit cell of AFM V$_2$WS$_4$ consists of two layers with AB stacking. The cyan arrow denotes for the half translation operator $\tau_{1/2}$. (b) The top view along the $z$ axis. (c) The monolayer atomic structure from top view. (d) The magnetic AFM-$z$ ground state. $J_{i}$ denote the leading magnetic couplings between V-V pairs, and negative $J_i$ means FM exchange coupling. $J_1=-13.0$~meV, $J_2=-7.5$~meV only represents intralayer next-nearest-neighbor V-V pairs through W atoms, $J_3=1.7$~meV and $J_4=1.4$~meV represent the two strongest interlayer couplings. (e) The calculated total energy for different magnetic ordered states.}
\label{fig1}
\end{figure}

\emph{Structure and magnetic properties.---}
The ternary transition metal chalcogenide V$_2$\textit{MX}$_4$, with $M=$ W or Mo, $X=$ S or Se, crystallize in an orthorhombic crystal structure with the space group $I\bar{4}2m$ (No.~121) with seven atoms in one primitive cell. Taking V$_2$WS$_4$ as an example, it has a layered structure with a tetragonal lattice V$_2$WS$_4$ as the building block shown in Fig.~\ref{fig1}. The key symmetry operation is $S_{4z}\equiv\mathcal{I}C_{4z}$, where $\mathcal{I}$ is inversion symmetry. Each layer contains three atomic sub-layers (i.e. one V$_2$W and two S$_2$), where each V or W atom is surrounded by four S atoms forming a distorted edge-sharing tetrahedron. The layers of bulk V$_2$WS$_4$ are connected by van der Waals interactions and stack along $z$ axis, forming an AB stacking pattern, which is energetically favorable than AA stacking~\cite{supple}. The B layer can be regarded as A layer translating along $\tau_{1/2}=(a,a,c)/2$, where $(a, c)=(5.79, 9.55)$~\r{A} are the in-plane and out-of plane lattice constant, respectively. The dynamical stability of V$_2$WS$_4$ is confirmed by first-principles phonon calculations~\cite{supple}. Moreover, Cu$_2\textit{MX}_4$ and Ag$_2\textit{MX}_4$ with the same structure have been successfully synthesized~\cite{crossland2005,hu2016,zhan2018,lin2019}, implying that these materials could potentially be fabricated experimentally.

\begin{figure*}
\begin{center}
\includegraphics[width=5.5in, clip=true]{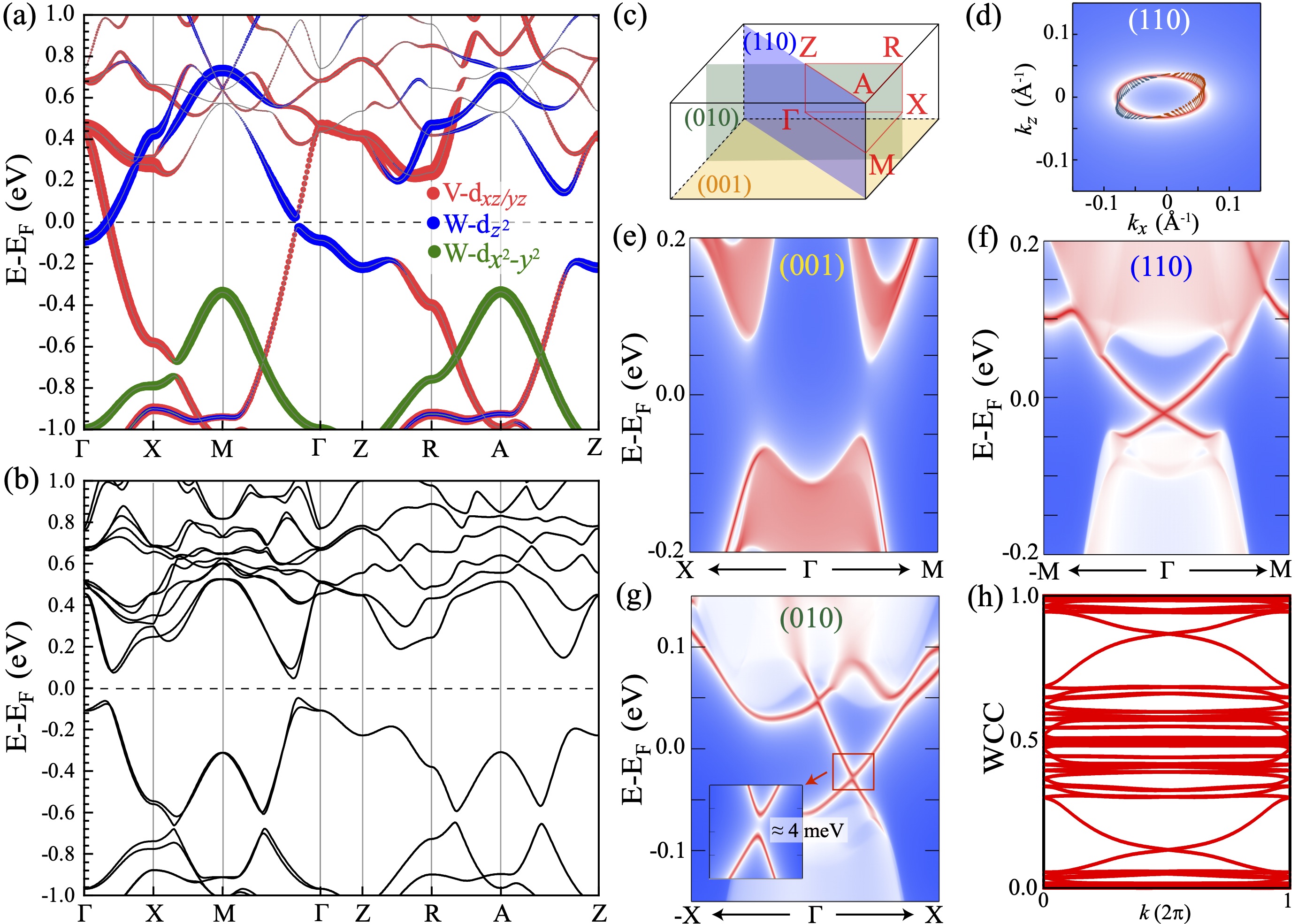}
\end{center}
\caption{Electronic structure and surfaces states of AFM-$z$ V$_2$WS$_4$. (a) The $d$-orbitals projected band structure without SOC, only those bands which are related to band inversion are highlighted. (b) Band structure with SOC. (c) Brillouin zone. $(110)$, $(001)$, and $(001)$ surfaces are labeled as blue, orange, and green, respectively. (d) Fermi surface with spin texture on the (110) surface at Fermi level presents an ellipse shape. Blue and orange arrow of spin texture denote $s_y<0$ and $s_y>0$, respectively. (e)-(g) Surface states of the semi-infinite (001), (110) and (010) surfaces, respectively, which is gapless on the $\mathcal{S}$-preserving (110) surface, but gapped on the $\mathcal{S}$-broken (001) and (010) surfaces. (h) The Wannier charge centers (WCC) is calculated in the plane including TRI momenta $(000)$, $(\pi\bar{\pi}0)$, $(\pi0\bar{\pi})$, and $(0\pi\bar{\pi})$ with $\mathbf{k}\cdot\tau_{1/2}=0$, confirming $\mathcal{Z}_2=1$.}
\label{fig2}
\end{figure*}

First-principles density functional theory (DFT) calculations are employed to investigate the electronic structure of V$_2$WS$_4$, with detailed methods provided in Supplemental Material~\cite{supple}. Our analysis reveals that each V atom has a valence of $+2$ by losing its two $4s$ electrons. Total energy calculations for various magnetic structures of 3D V$_2$WS$_4$ were performed, as summarized in Fig.~\ref{fig1}(e). The results show that the $A$-type AFM state with an out-of-plane easy axis (denoted as AFM-$z$) is the magnetic ground state. In this configuration, the material is ferromagnetic (FM) within the $xy$ plane of each layer and AFM between adjacent layers along the $z$ direction [Fig~\ref{fig1}(d)]. The total energy of $A$-type AFM state with in-plane easy axis (AFM-$x$) is slightly higher than that of AFM-$z$, but significantly lower than FM-$z$ state with an out-of-plane easy axis. This indicates that the magnetic anisotropy energy is weaker than the effective magnetic exchange interaction between neighboring layers. The calculated magnetic moments are primarily contributed by V ($\approx 2.6\mu_{B}$), with a smaller contribution from W ($\approx 0.4\mu_{B}$), confirming that the magnetism originates from the V atoms. This fractional magnetic moment arises from the band inversion between V $d_{xz,yz}$ orbitals and W $d_{z^2}$ orbital [Fig.~\ref{fig2}(a)]. The tetrahedral crystal field splits V $3d$ orbitals into lower-energy $e_g(d_{z^2},d_{x^2-y^2})$ and higher-energy triplet $t_{2g}(d_{xz/yz},d_{xy})$. The three remaining $3d$ electrons occupy the spin-up V-$d$ levels, forming an $e_g^2t_{2g}^1$ configuration with a magnetic moment of approximately $3\mu_B$ according to the Hund’s rule, which is close to the DFT calculation. The FM exchange coupling between neighboring V atoms within each layer is strongly enhanced by Hund's rule interaction due to empty $t_{2g}$ orbitals~\cite{khomskii2004}.  The $t_{2g}$-$t_{2g}$ superexchange of V atoms between adjacent layers via $p$ orbitals of ligand is AFM due to the Goodenough-Kanamori-Anderson rule~\cite{khomskii2004}. Futhermore, the N\'{e}el temperature for AFM-$z$ V$_2$WS$_4$ is estimated as $490$~K by Monte Carlo simulations~\cite{supple}.

\emph{AFM TI and topological invariant.---} First we investigate the AFM-$z$ ground state, which belongs to the type IV magnetic space group No.~114.282 in Belov-Neronova-Smirnova (BNS) notation~\cite{bradley1972}. The symmetry generators of this group include $S_{4z}$, $\Theta\tau_{1/2}$, $C_{2z}$ and $C_{2x}\tau_{1/2}$. While the time-reversal symmetry $\Theta$ is broken, a combined symmetry $\mathcal{S}\equiv\Theta\tau_{1/2}$ is preserved, where $\tau_{1/2}$ is the half translation operator connecting neighboring W atomic layers, as marked in Fig.~\ref{fig1}(a). This combined symmetry $\mathcal{S}$ is antiunitary and satisfies $\mathcal{S}^2=-e^{-2i\mathbf{k}\cdot\tau_{1/2}}$. On Brillouin-zone (BZ) plane where $\mathbf{k}\cdot\tau_{1/2}=0$, $\mathcal{S}^2=-1$ . Therefore, similar to $\Theta$ in time-reversal-invariant (TRI) TI, $\mathcal{S}$ enables a $\mathcal{Z}_2$ classification~\cite{mong2010}, where the $\mathcal{Z}_2$ topological invariant is well defined on the BZ plane with $\mathbf{k}\cdot\tau_{1/2}=0$. The electronic structure without and with spin-orbital coupling (SOC) are shown in Fig.~\ref{fig2}(a) and Fig.~\ref{fig2}(b), respectively. One can see an anticrossing feature around $\Gamma$ point from the band inversion between V $d_{xz,yz}$ orbitals and W $d_{z^2}$ orbital, suggesting that V$_2$WS$_4$ might be topologically nontrivial. Since $\mathcal{I}$ is broken but $S_{4z}$ is preserved, the $\mathcal{Z}_2$ invariant is simply determined by the $S_{4z}$ eigenvalues of the wavefunctions at $S_{4z}$-invariant momenta in the BZ~\cite{elcoro2021}, with the explicit form
\begin{equation}
\mathcal{Z}_2=\sum_{K=\Gamma,M,Z,A}\frac{1}{2}\left(n_K^{\frac{1}{2}}-n_K^{-\frac{3}{2}}\right)\mathrm{mod}~2,
\end{equation}
where $n_K^{1/2}$ and $n_K^{-3/2}$ are the number of occupied states with $S_{4z}$ eigenvalues $e^{-i\pi/4}$ and $e^{i3\pi/4}$, respectively. $K=\Gamma,M,Z,A$ are $S_{4z}$ invariant in BZ. $n_K^{1/2}$ and $n_K^{-3/2}$ of high symmetry points are listed in Table~\ref{tab1}, so $\mathcal{Z}_2=1$. There are two additional symmetry indicators $\mathcal{Z}_4$ and $\delta_2$, which are used to characterize higher-order topology and Weyl semimetal~\cite{song2018,khalaf2018}. In the case of V$_2$WS$_4$, both $\mathcal{Z}_4=\delta_2=0$, as the Chern number for all $k_z$ planes in the BZ is consistently zero. The monolayer is a FM QAH insulator~\cite{jiang2024}, thus 3D AFM-$z$ V$_2$WS$_4$ can be viewed as successive stacking of layered QAH with alternating Chern number $\mathcal{C}=\pm1$, which are related by $\mathcal{S}$ symmetry. We further employ the Willson loop method~\cite{yu2011} to confirm the $\mathcal{Z}_2$ topological invariant in Fig.~\ref{fig2}(h), concluding that AFM-$z$ V$_2$WS$_4$ is indeed an AFM TI. Notably, we notice that a large energy gap of about $0.1$~eV is obtained in Fig.~\ref{fig2}(b).

\begin{table}[t]
\caption{The number of occupied bands with $S_{4z}$ eigenvalue $e^{-i\pi/4}$ and $e^{i3\pi/4}$ at four high symmetry points in BZ.}
\begin{center}\label{tab1}
\renewcommand{\arraystretch}{1.5}
\begin{tabular*}{3.4in}
{@{\extracolsep{\fill}}ccccc}
\hline
\hline
$K$ & $\Gamma (000)$ & $M (\pi\pi0)$ & $Z (00\pi)$ & $A (\pi\pi\pi)$
\\
\hline
$n_K^{1/2},n_K^{-3/2}$  & $20,20$ & $20,20$ & $20,20$ & $19,21$
\\
\hline
\hline
\end{tabular*}
\end{center}
\end{table}

One prominent feature of AFM TI is the existence of gapless surface states that depends on the crystallographic orientation of the surface plane, which is confirmed by the surface-state calculations. As shown in Fig.~\ref{fig2}(f), the gapless surface states can be seen at $\Gamma$ point forming a single Dirac cone in bulk gap on $\mathcal{S}$-preserving (110) surface. While the surface states are gapped on $\mathcal{S}$-broken (001) and (010) surfaces, as shown in Fig.~\ref{fig2}(e) and Fig.~\ref{fig2}(g), respectively.

\emph{Axion state and TME.---}
The $\mathcal{Z}_2=1$ topological invariant of AFM-$z$ V$_2$WS$_4$ with a full band gap signifies the axion state with a quantized value $\theta=\pi$~(mod~$2\pi$), where the electromagnetic response is described by the axion electrodynamics, $S_{\theta}=(\theta/2\pi)(\alpha/2\pi)\int d^3 xdt \mathbf{E}\cdot \mathbf{B}$. Here, $\mathbf{E}$ and $\mathbf{B}$ are the conventional electromagnetic fields inside the insulator, $\alpha=e^2/\hbar c$ is the fine-structure constant, $e$ is electron charge, and $\theta$ is dimensionless pseudoscalar parameter~\cite{qi2008}. This axion state gives rise to the TME, a phenomenon yet to be observed experimentally~\cite{tokura2019}. Interestingly, the gapped surface states from time-reversal symmetry breaking are naturally and intrinsically provided by even-layer V$_2$WS$_4$ films with $A$-type AFM structure, which make it an ideal platform for the long-sought quantized TME. Furthermore, to observe TME, all surface states must be gapped~\cite{wang2015b}, which could be fulfilled by synthesizing realistic materials without any $S$-preserving surfaces. Compared to the previous proposals on TME in FM-TI heterostructure~\cite{qi2008,wang2015b,morimoto2015}, the intrinsic magnetic TI material V$_2$WS$_4$ offer a more practical and promising avenue for exploring axion electrodynamics.

\begin{figure}[t]
\begin{center}
\includegraphics[width=3.4in, clip=true]{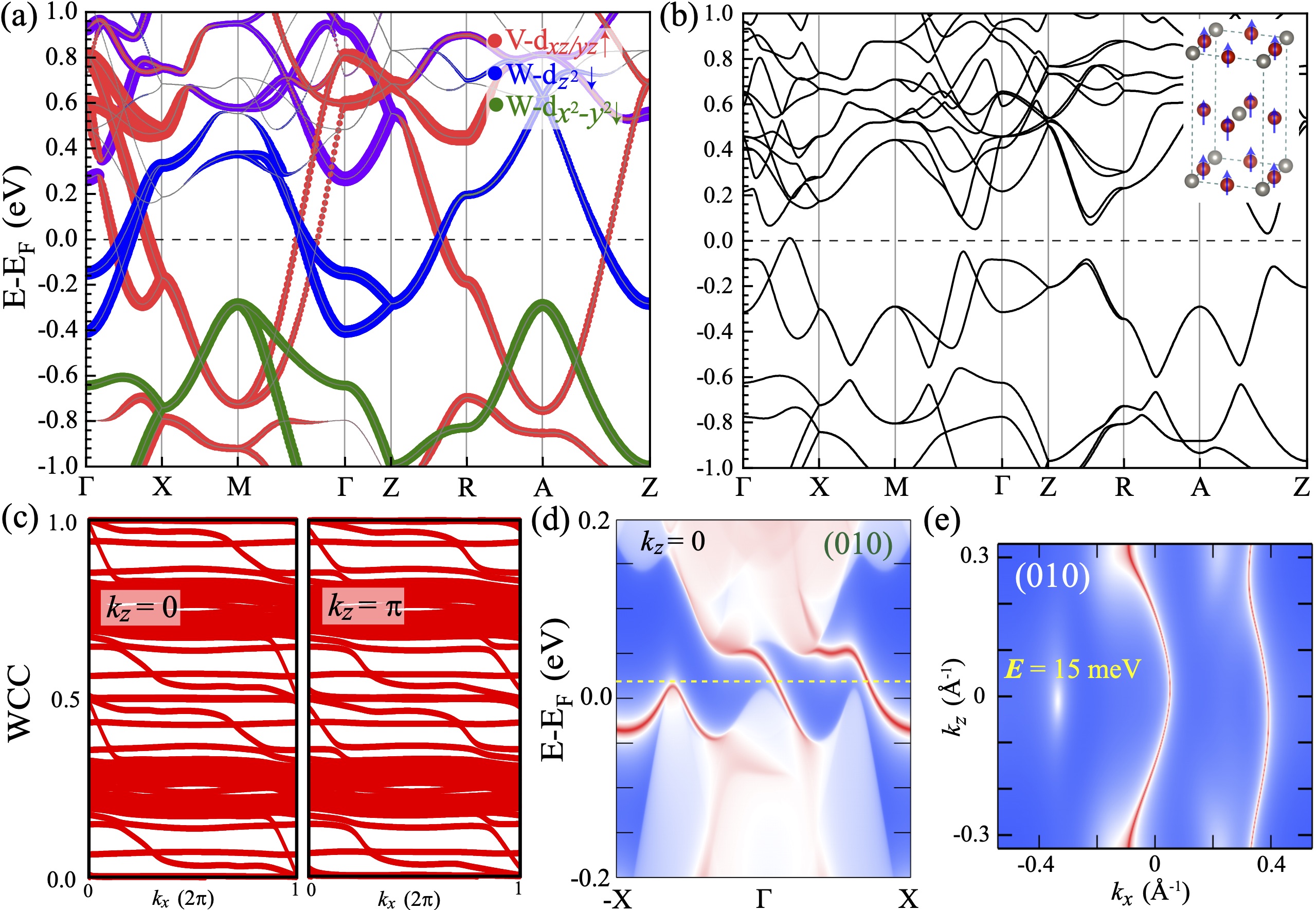}
\end{center}
\caption{Electronic structure and surfaces states of FM-$z$ V$_2$WS$_4$. (a) The $d$-orbital projected band structure without SOC. (b) Band structure with SOC. Here the conventional unit cell with FM-$z$ state is chosen. (c) The evolution of WCC along the $k_x$ direction in the $k_z=0$ and $k_z=\pi$ plane. (d) Two chiral surface states on the semi-infinite (010) surface at $k_z=0$. Consistent with WCC in (c), the Chern number is $\mathcal{C}=-2$ from $k_z=0$ to $k_z=\pi$. (e) Fermi surface on the (010) termination at the isoenergy $15$~meV above Fermi level. The chiral surface states extend over the entire surface BZ from $k_z=-\pi$ to $k_z=\pi$, indicating the 3D QAH state.}
\label{fig3}
\end{figure}

\emph{3D QAH state.---}
The AFM ground state of V$_2$WS$_4$ could be tuned to a FM configuration by applying an external magnetic field, leading to distinct topological phases. Here we study FM-$z$ state, which belongs to magnetic space group No.~121.331 in BNS notation with symmetry generators $S_{4z}$, $C_{2z}$ and $\Theta C_{2y}$. For simplicity, we adopted the conventional unit cell here as in AFM-$z$ state. Then 3D FM-$z$ V$_2$WS$_4$ can be interpreted as layer stacking of FM QAH insulator with the same Chern number $\mathcal{C}=-1$, leading to 3D QAH state or Weyl semimetal~\cite{burkov2011,wang2016a}. The electronic structures without and with SOC are calculated in Fig.~\ref{fig3}(a) and Fig.~\ref{fig3}(b), respectively. There is spin polarized band inversion near the Fermi energy between spin up $d_{xz},d_{yz}$ bands of V and spin down $d_{z^2}$ band of W, which is further gapped by SOC. Interestingly, along $\Gamma$-$Z$ line, the band inversion always remains and there is no level crossing. Meanwhile, the FM-$z$ state remains insulating, with a gap of approximately 20 meV. The Willson loop calculations shown in Fig.~\ref{fig3}(c) reveal the Chern number $\mathcal{C}=-2$ at both $k_z=0$ and $k_z=\pi$ planes, confirming that the system is a 3D QAH state. Moreover, our surface state calculations demonstrated the existence of chiral surface state on the (010) termination, which is the fingerprint of 3D QAH state. As shown in Fig.~\ref{fig3}(d), two chiral edge states disperse within the bulk gap at $k_z=0$ plane. Such chiral edge states extend over the entire surface BZ from $k_z=-\pi$ to $k_z=\pi$ plane without any Weyl points [Fig.~\ref{fig3}(e)]. Thin films of 3D QAH insulator lead to the QAH effect in two dimensions (2D), the Chern number of which is equal to the layer number as will be discussed later. The high Chern number QAH effect with multiple dissipationless edge channels could lead to novel design of low energy cost electronic devices.

\emph{Tight-binding model and multilayer.---}
The layered van der Waals materials are featured by tunable quantum size effects. Here the band inversion in 3D suggests nontrivial topology may also exist in 2D multilayers. For AFM V$_2$WS$_4$ films, even and odd layers have distinct symmetry and topological properties. Even layers have $S_{4z}$ and $C_{2x}\tau'_{1/2}$ symmetries, where $\tau'_{1/2}\equiv(a,a,0)/2$, and all of the bands have Chern number $\mathcal{C}=0$ because of the Hall conductance $\sigma_{xy}$ is odd under $C_{2x}\tau'_{1/2}$. Differently, odd layers have $S_{4z}$ and $C_{2x}\Theta$ symmetries, $\mathcal{C}\neq0$ is allowed for $\sigma_{xy}$ is invariant under $C_{2x}\Theta$. We construct a tight-binding model to recover the essential topological physics for AFM ground state, and investigate the crossover between bulk and multilayers.

From DFT calculations in Fig.~\ref{fig2}, we construct the minimal tight-binding model including $d_{xz},d_{yz}$ of V and $d_{z^2},d_{x^2-y^2}$ of W, where the band gap is mainly provided by the intralayer SOC effect with the opposite spin and interlayer orbital hopping with the same spin. The Hamiltonian is written as $\mathcal{H}=\begin{pmatrix} \mathcal{H}_1 & \mathcal{T} \\\mathcal{T}^{\dag} & \mathcal{H}_{2} \\ \end{pmatrix}$, where $\mathcal{H}_{1,2}$ are the intralayer Hamiltonian for two layers in the unit cell, $\mathcal{T}$ is the interlayer hopping. For the intralayer, there are two V atoms, and $d_{xz},d_{yz}$ orbitals of each V are non-degenerate. However, $d_{xz}$ of one V and $d_{yz}$ of the other V are degenerate, which are related to each other by $S_{4z}$. Therefore, for the low-energy physics of intralayer, for instance $\mathcal{H}_1$, we only need to consider $d^\uparrow_{1,xz},d^\uparrow_{1,yz}$ from two V, respectively and $d_{1,z^2}^\downarrow, d_{1,x^2-y^2}^\downarrow$ of W, namely a four orbitals model. $\mathcal{H}_1$ is obtained by considering the nearest-neighbor and next-nearest-neighbor hopping with SOC included. Then $\mathcal{H}_2$ of other layer is related to $\mathcal{H}_1$ by $\mathcal{S}$ symmetry, where the spins are flipped with the basis of $d_2\equiv$ ($d^\downarrow_{2,xz}$,$d^\downarrow_{2,yz}$,$d^\uparrow_{2,z^2}$,$d^\uparrow_{2,x^2-y^2}$)$^T$. The interlayer hopping $\mathcal{T}$ includes the orbital overlapping with the same spin, with the strength of about $50$~meV which is smaller than the intralayer SOC. The explicit forms and fitted parameters are listed in Supplemental Material, where similar electronic structure and surface states of our model are obtained as DFT calculations~\cite{supple}.

\begin{figure}[t]
\begin{center}
\includegraphics[width=3.3in, clip=true]{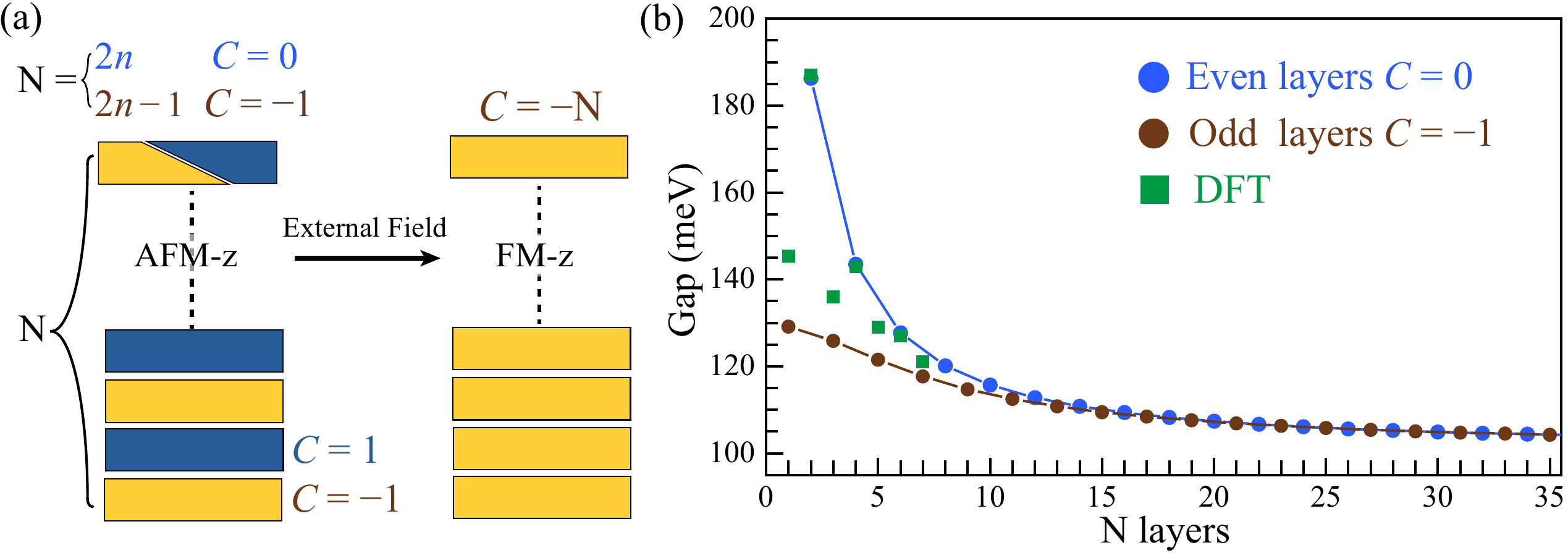}
\end{center}
\caption{(a) Schematic diagram shows layered stacking of QAH insulators with alternating Chern number $\mathcal{C}=\pm1$. In AFM-$z$ state, it gives $\mathcal{C}=-1$ QAH insulator in odd layers and $C=0$ axion insulator in even layers. In FM-$z$ state viewed as stacking of same Chern number $\mathcal{C}=-1$ QAH insulators, $N$ layers is a $\mathcal{C}=-N$ high Chern number QAH insulator. (b) The evolution of band gap as a function of layer number in AFM-$z$ state. Blue (brown) circle denotes even (odd) layers calculated from tight-binding model. Green box represents the gaps of one to seven layers by DFT calculations~\cite{supple}.}
\label{fig4}
\end{figure}

By utilizing the tight-binding model, we can study the dimensional crossover from bulk to multilayer. As shown in Fig.~\ref{fig4}, in AFM-$z$ ground state, the band gap of multilayer shows oscillatory decay behavior and gradually converges to bulk value when the film exceed twenty layers, while the Chern number exhibit pronounced even-odd oscillations. The Chern number of band in a $S_{4z}$ invariant system is $i^\mathcal{C}=\prod_{j\in\text{occupied}} (-1)^F \xi_j(\Gamma)\xi_j(M)\zeta_j(X)$, with $F=1$ for spinful case here, $\xi_j(k)$ is the $S_{4z}$ eigenvalue at $\Gamma$ and M points of the $j$-th band, $\zeta_j(X)$ is the $C_{2z}$ eigenvalue at X point on the $j$-th band~\cite{jiang2024}. Explicitly, odd layers have $\mathcal{C}=-1$, while even layers have $\mathcal{C}=0$. Our DFT calculations up to seven layers are consistent with effective model~\cite{supple}. These results suggest that multilayer V$_2$WS$_4$ can be viewed as layered stacking of alternating $\mathcal{C}=\pm1$ QAH insulators for AFM-$z$ state, or stacking of same $\mathcal{C}=-1$ QAH insulators for FM-$z$ state, as illustrated in Fig.~\ref{fig4}(a). The interlayer coupling is weaker than band inversion and SOC, thus the Chern number of multilayer is simply the summation of Chern number from each layer, namely
\begin{equation}
\mathcal{C}_{\text{multilayer}}=\sum\limits_{j}\mathcal{C}_j.
\end{equation}
Here $\mathcal{C}_j=\pm1$ for each layer is only determined by the direction of magnetic moment, and does not affected by interlayer coupling.

It is insightful to compare V$_2$WS$_4$ with MnBi$_2$Te$_4$, as both materials are layered van der Waals intrinsic magnetic TI with similar topological properties. In their AFM-$z$ ground state, both are classified as AFM TIs and axion insulators in 3D, displaying an oscillation between zero and odd Chern numbers in AFM multilayers.  
However, their phases diverge significantly in the FM-$z$ state.
While V$_2$WS$_4$ is a 3D QAH state, MnBi$_2$Te$_4$ tends to be a Weyl semimetal or a trivial FM insulator~\cite{zhang2019,li2019}. 
The primary distinctions arise from differences in interlayer coupling and band inversion. First, the low-energy physics in V$_2$WS$_4$ is from the $d$-orbitals of V and W, which are located in the middle atomic layer. This contrasts with MnBi$_2$Te$_4$, where the low-energy physics is dominated by the $p_z$ orbitals of the outermost Bi/Te atomic layers. Consequently, the interlayer coupling is indirect and much weaker in V$_2$WS$_4$ compared to MnBi$_2$Te$_4$ (about $0.1$~eV). Second, V$_2$WS$_4$ possesses much deeper band inversion. The band inversion point lie approximately at 35\% along $\Gamma$-$X$ line from $\Gamma$ ($|\mathbf{k}|=0.19$ \r{A}$^{-1}$) as shown in Fig.~\ref{fig2}(a), which is further opened by a strong SOC. The weak interlayer coupling could not change the band inversion along $\Gamma$-$Z$ in V$_2$WS$_4$. In contrast, the SOC induced band inversion is at $\Gamma$ in MnBi$_2$Te$_4$, then a relatively stronger interlayer coupling could modify the band inversion and lead to trivial insulators in few layers and Weyl semimetal in 3D. These distinctions highlight a key insight that such characteristics do not arise from fortuitousness in parameters, but rather from the universality inherent in V$_2$WS$_4$ family.

\emph{Discussion.---}
Other ternary transition metal chalcogenide, such as V$_2$WSe$_4$, V$_2$Mo\textit{X}$_4$ and Ti$_2$W\textit{X}$_4$ ($X=$ S or Se), which share the same orthorhombic crystal structure, are also promising candidates for hosting magnetic topological states similar to V$_2$WS$_4$. In fact, most of them are found to be AFM TI in the ground state, as calculated in the Supplemental Material~\cite{supple}. The synergy between intrinsic magnetism and topologically nontrivial bands, along with the variety of candidate materials, provides a rich platform for exploring emergent phenomena in magnetic topological states across different spatial dimensions. For instance, the magnetic fluctuations in these systems also give dynamic axion field.

The field of topological quantum matter in recent years developed explosively in materials science and condensed matter physics. One of main reasons is the precise theoretical predictions and experimental discovery of intrinsic topological materials. Tracing back the research history in magnetic topological physics, most of the previous experimental works are based on magnetically doped TIs and heterostructures~\cite{chang2013b,checkelsky2014,kou2014,bestwick2015,chang2015,mogi2017,mogi2017a,grauer2017,xiao2018}, which are quite complex and challenge to study~\cite{chong2020,lachman2017}. The research progress have been greatly prompted by discovering intrinsic magnetic TI material MnBi$_2$Te$_4$~\cite{deng2020,liu2020}. However, the co-antisite defects in Mn and Bi layers drastically suppress the exchange gap by several order of magnitude~\cite{garnica2022,tan2023,wu2023}, which fundamentally deteriorates magnetic topological states. Meanwhile, few layers MnBi$_2$Te$_4$ with topologically nontrivial bands are too thick to tune efficiently. Finally, layered MnBi$_2$Te$_4$ usually contain Bi$_2$Te$_3$ layers, which further complicates the electronic structure with undesired topology. The V$_2$WS$_4$-family materials satisfy all these material characteristics of simple, magnetic and topological. For example, monolayer V$_2$WS$_4$ is QAH insulator, in contrast to trivial FM insulator of monolayer MnBi$_2$Te$_4$. Therefore, the techniques developed for 2D materials with versatile tunability can be readily applied to V$_2$WS$_4$ family. We anticipate that van der Waals heterostructures integrating V$_2$WS$_4$ family with other magnetic or superconducting 2D materials will provide fertile ground for exploring exotic topological quantum phenomena.

In summary, our work uncovers a large class of intrinsic magnetic TI materials with extremely rich topological quantum states of exceptional characteristics in different spatial dimensions. The broad range of candidate materials suggests that the underlying physics is quite general. We anticipate this will further enrich the magnetic TI family and provide a new material platform for exotic topological phenomena.

\begin{acknowledgments}
This work is supported by the Natural Science Foundation of China through Grants No.~12350404 and No.~12174066, the Innovation Program for Quantum Science and Technology through Grant No.~2021ZD0302600, the Science and Technology Commission of Shanghai Municipality under Grants No.~23JC1400600, No.~24LZ1400100 and No.~2019SHZDZX01. Y.J. acknowledges the support from China Postdoctoral Science Foundation under Grants No.~GZC20240302 and No.~2024M760488.
\end{acknowledgments}

\end{document}